# Effect of Top Al$_2$O$_3$ Interlayer Thickness on Memory Window and Reliability of FeFETs With TiN/Al$_2$O$_3$/Hf$_{0.5}$Zr$_{0.5}$O$_2$/SiO$_x$/Si (MIFIS) Gate Structure

Tao Hu, Xinpei Jia, Runhao Han, Jia Yang, Mingkai Bai, Saifei Dai, Zeqi Chen, Yajing Ding, Shuai Yang, Kai Han, Yanrong Wang, Jing Zhang, Yuanyuan Zhao, Xiaoyu Ke, Xiaoqing Sun, Junshuai Chai, Hao Xu, Xiaolei Wang, Wenwu Wang and Tianchun Ye

*Abstract*—We investigate the effect of top Al$_2$O$_3$ interlayer thickness on the memory window (MW) of Si channel ferroelectric field-effect transistors (Si-FeFETs) with TiN/Al$_2$O$_3$/Hf$_{0.5}$Zr$_{0.5}$O$_2$/SiO$_x$/Si (MIFIS) gate structure. We find that the MW first increases and then remains almost constant with the increasing thickness of the top Al$_2$O$_3$. The phenomenon is attributed to the lower electric field of the ferroelectric Hf$_{0.5}$Zr$_{0.5}$O$_2$ in the MIFIS structure with a thicker top Al$_2$O$_3$ after a program operation. The lower electric field makes the charges trapped at the top Al$_2$O$_3$/Hf$_{0.5}$Zr$_{0.5}$O$_2$ interface, which are injected from the metal gate, cannot be retained. Furthermore, we study the effect of the top Al$_2$O$_3$ interlayer thickness on the reliability (endurance characteristics and retention characteristics). We find that the MIFIS structure with a thicker top Al$_2$O$_3$ interlayer has poorer retention and endurance characteristics. Our work is helpful in deeply understanding the effect of top interlayer thickness on the MW and reliability of Si-FeFETs with MIFIS gate stacks.

This work was supported in part by the National Natural Science Foundation of China under Grant Nos. 92264104 and 52350195, in part by National Key Research and Development Program of China under Grant No. 2022YFB4400300, in part by R&D Program of Beijing Municipal Education Commission under Grant No.KZ202210009014, in part by the Young Elite Scientists Sponsorship Program under Grant No. BYESS2023033, and Supported by the Postdoctoral Fellowship Program of CPSF under Grant No. GZC20232925. (Corresponding author: Xiaolei Wang, Hao Xu)

Tao Hu, Xinpei Jia, Runhao Han, Jia Yang, Mingkai Bai, Saifei Dai, Zeqi Chen, Yajing Ding, Shuai Yang, Yuanyuan Zhao, Xiaoyu Ke, Xiaoqing Sun, Junshuai Chai, Hao Xu, Xiaolei Wang, Wenwu Wang and Tianchun Ye are with Institute of Microelectronics of the Chinese Academy of Sciences, Beijing 100029, China. The authors are also with the School of Integrated Circuits, University of Chinese Academy of Sciences, Beijing 100049, China (e-mail: wangxiaolei@ime.ac.cn, xuhao@ime.ac.cn).

Yanrong Wang and Jing Zhang are with School of Information Science and Technology, North China University of Technology, Beijing 100144, China.

Kai Han is with the School of Physics and Electronic Information, Weifang University, Weifang, Shangdong 261061, China.

## I. INTRODUCTION

The massive data storage market is dominated by NAND flash memory. The primary drivers of NAND flash technology development are density and cost per bit [1]. To meet the growing demand for high-density storage, 3D vertical NAND (3D VNAND) has replaced 2D NAND [2]. However, due to the strong cell-to-cell interference under the high operation voltages, 3D VNAND currently faces scaling limits in the Z-direction [1, 3, 4]. Ferroelectric-based vertical NAND (Fe-VNAND) is one of the strong candidates for 3D VNAND technology to overcome the scaling limits [1, 3-7]. However, Fe-VNAND faces the huge challenge of a narrow memory window (MW) for the application in high-density storage. Generally, hafnia-based silicon channel ferroelectric field-effect transistors (HfO$_2$-based Si-FeFETs) with metal/HfO$_2$-based ferroelectric/bottom interlayer/silicon substrate (MFIS) gate structure has a MW of about 1-2 V [8-11]. The physical origin of narrow MW are low coercive voltage ($V_c$) of ferroelectric Hf$_{0.5}$Zr$_{0.5}$O$_2$ within the thickness range of 20 nm, and a significant charge trapping effect between the silicon channel and ferroelectric Hf$_{0.5}$Zr$_{0.5}$O$_2$/SiO$_x$ interface due to the presence of the large spontaneous polarization ($P_s$) of ferroelectric doped-HfO$_2$ (~ 20-30 μC/cm$^2$) [12-15]. However, the coercive field of the ferroelectric Hf$_{0.5}$Zr$_{0.5}$O$_2$ is difficult to change on a large range. Therefore, to enlarge the MW, several studies focus on suppressing the charge injection from the silicon channel to the ferroelectric Hf$_{0.5}$Zr$_{0.5}$O$_2$/SiO$_x$ interface, such as reducing the spontaneous polarization of the ferroelectric [16-19], eliminating the bottom interlayer [20-24], and applying high-κ interlayer [25-29]. However, the above methods still cannot significantly enlarge the MW. This small MW does not meet the requirements of the multi-bit memory cell. Recently, some studies have found that inserting a dielectric interlayer (e.g. Al$_2$O$_3$ or SiO$_2$) between the metal gate and the ferroelectric layer can significantly improve the MW of Si-FeFETs [1, 3, 4, 24, 30-35]. The MW can achieve 6.4 V (or 8.3 V) by inserting 3.4 nm (or 4 nm) SiO$_2$ and 4.1 V by inserting 3 nm Al$_2$O$_3$ [32, 33, 35]. Moreover, the simulation

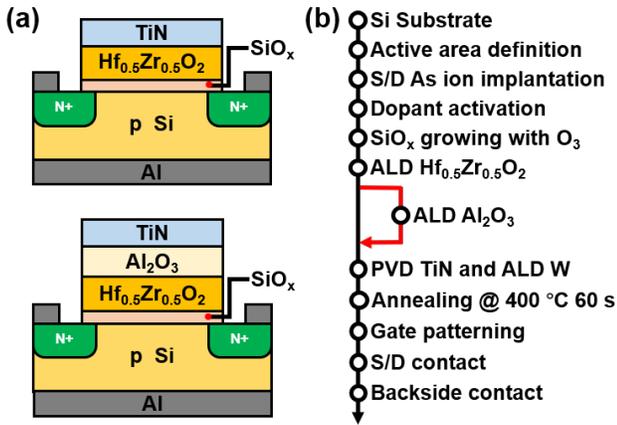

Fig. 1. (a) Schematic of the HfO$_2$-based Si-FeFETs device structure and (b) fabrication process flow.

results from [1] show that using a thicker top interlayer is beneficial for the MW increase.

However, there are still few experimental studies on the effect of the top Al$_2$O$_3$ interlayer thickness on the MW and reliability of Si-FeFETs. Therefore, we experimentally study and report the dependence of the MW and reliability on the top Al$_2$O$_3$ interlayer thickness and discuss its physical origin in this work. We find that the MW first increases and then remains almost constant with the increasing thickness of the top Al$_2$O$_3$. The phenomenon is attributed to the lower electric field of the ferroelectric Hf$_{0.5}$Zr$_{0.5}$O$_2$ in the MIFIS structure with a thicker top Al$_2$O$_3$ after a program operation. The lower electric field makes the charges trapped at the top Al$_2$O$_3$/Hf$_{0.5}$Zr$_{0.5}$O$_2$ interface ($Q_{it'}$), which are injected from the metal gate, cannot be retained. Furthermore, we study the effect of the top Al$_2$O$_3$ interlayer thickness on the reliability (endurance characteristics and retention characteristics). We find that the MIFIS structure with a thicker top Al$_2$O$_3$ interlayer has poorer retention characteristics and endurance characteristics.

## II. DEVICE FABRICATION AND CHARACTERIZATION

Fig. 1 shows the schematic of the HfO$_2$-based Si-FeFETs device structure and fabrication process flow. In our work, there are two different gate stacks. One is TiN/Hf$_{0.5}$Zr$_{0.5}$O$_2$/SiO$_x$/Si (MFIS) as the control sample. The other is TiN/Al$_2$O$_3$/Hf$_{0.5}$Zr$_{0.5}$O$_2$/SiO$_x$/Si (MIFIS) with a 0.85, 1.7, 2.55, 4.5, 5.5, 8, or 13 nm top Al$_2$O$_3$ dielectric interlayer. The detailed fabrication procedure of these devices can be found in the previous report [32].

Fig. 2 shows High-Resolution Transmission Electron Microscopy (HRTEM) images and Energy Dispersion Spectrometer (EDS) results for both MFIS and MIFIS structures, where the MIFIS structure with 4.5 nm top Al$_2$O$_3$ interlayer is shown as an example of the MIFIS structure with different top Al$_2$O$_3$ thicknesses. For the MIFIS structure, the presence of a peak concentration of Al at the TiN/Hf$_{0.5}$Zr$_{0.5}$O$_2$ interface confirms the presence of the top Al$_2$O$_3$ interlayer.

The gate length/width (L/W) of these devices in this work is 5/150 µm. The electrical measurements were performed by Keysight B1500A with waveform generator fast measurement unit (WGFMU) and high voltage semiconductor pulse generator unit (SPGU). Fig. 3(a) shows the schematic of the electrical measurement of these devices. Fig. 3(b-d) shows the measurement waveforms of MW, endurance characteristics, and retention characteristics. The threshold voltage ($V_{th}$) is extracted by the constant current method at the drain current $I_d$ = W/L × 10$^{-7}$ A.

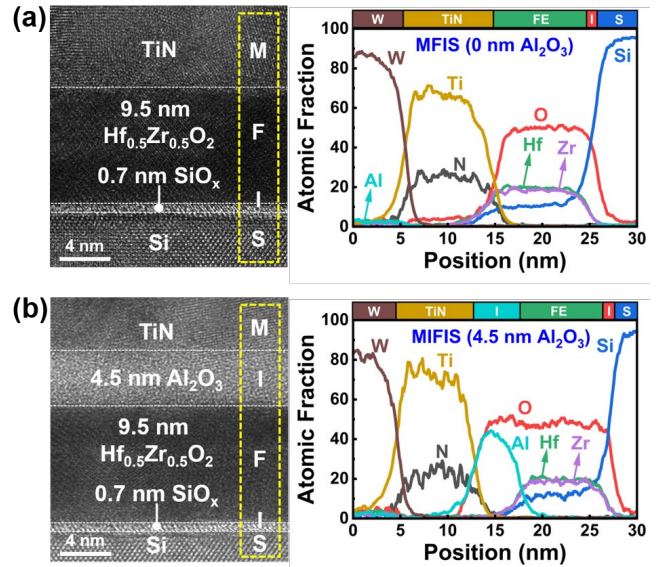

Fig. 2. HRTEM images and EDS of the (a) MFIS and (b) MIFIS structures with 4.5 nm top Al$_2$O$_3$ interlayer.

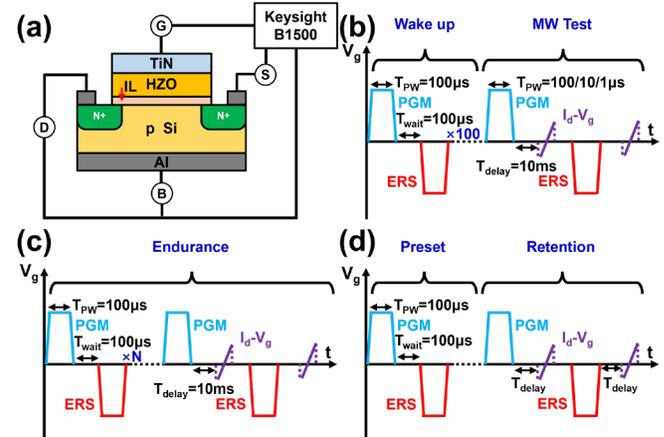

Fig. 3. (a) Schematic of electrical measurement of these devices. The measurement waveforms of (b) MW, (c) endurance characteristics and (d) retention characteristics.

## III. MODEL

For the MIFIS gate structure as shown in Fig. 1(a), the voltage distribution across the gate stacks is given as

$$V_g = \varphi_s + V_{FE} + V_{BIL} + V_{TIL} \qquad (1)$$

where $V_g$ is the gate voltage, $\varphi_s$, $V_{FE}$, $V_{BIL}$, and $V_{TIL}$ are the voltage drop across Si substrate, ferroelectric Hf$_{0.5}$Zr$_{0.5}$O$_2$, bottom SiO$_x$ interlayer, and top Al$_2$O$_3$ interlayer, respectively. The work function difference between the metal gate and the Si substrate is set to zero here. According to the charge neutrality condition, we can obtain

$$Q_M + Q_{it'} + Q_{it} + Q_s = 0 \qquad (2)$$

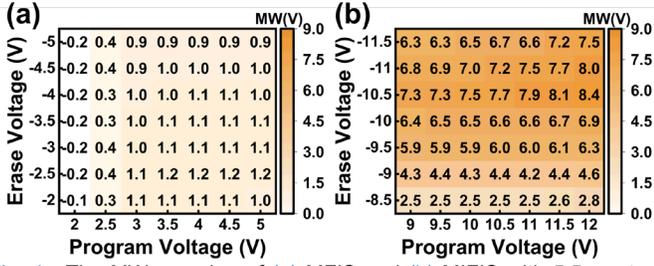

Fig. 4. The MW mapping of (a) MFIS and (b) MIFIS with 5.5 nm top Al$_2$O$_3$ as a function of the pulse amplitude under the pulse width of 100 µs.

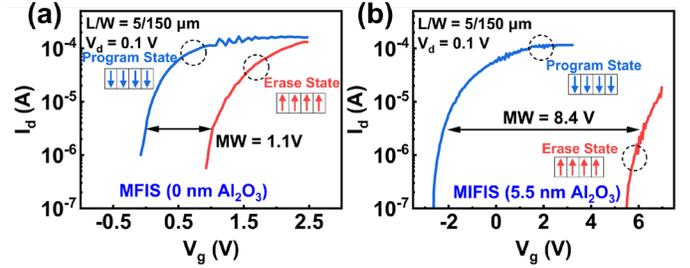

Fig. 5. $I_d$–$V_g$ curves of the maximum MW for (a) MFIS and (b) MIFIS with 5.5 nm top Al$_2$O$_3$ interlayer under the pulse width of 100µs.

where $Q_M$, $Q_{it}$, $Q_{it'}$, and $Q_s$ are the charge density of the metal gate, bottom SiO$_x$/Hf$_{0.5}$Zr$_{0.5}$O$_2$ interface, top Al$_2$O$_3$/Hf$_{0.5}$Zr$_{0.5}$O$_2$ interface, and Si substrate, respectively. $V_{TIL}$ is given as

$$V_{TIL} = \frac{Q_M}{C_{TIL}} \quad (3)$$

where $C_{TIL}$ is the capacitance of the top Al$_2$O$_3$ interlayer. $V_{BIL}$ is given as

$$V_{BIL} = \frac{-Q_{Si}}{C_{BIL}} = \frac{Q_M + Q_{it'} + Q_{it}}{C_{BIL}} \quad (4)$$

where $C_{BIL}$ represents the capacitance of the bottom SiO$_x$ interlayer. The relationship between the charge density of the Si substrate $Q_s$ and the Si surface potential $\varphi_s$ is given by [11]

$$Q_s(\varphi_s) = -SGN(\varphi_S)\sqrt{2}(\varepsilon_0\varepsilon_s / \beta L_D) \\ \times ((e^{-\beta\varphi_s} + \beta\varphi_s - 1) + (\frac{n_i}{N_a})^2(e^{\beta\varphi_s} - \beta\varphi_s - 1))^{1/2} \quad (5)$$

where $\beta = q/(kT)$, $L_D = (\varepsilon_0\varepsilon_s/q\beta N_a)$, $\varepsilon_0$ is the vacuum dielectric constant, $\varepsilon_s$ is the relative dielectric constant of the silicon, $n_i$ is the intrinsic carrier concentration of the silicon, and $N_a$ is the substrate doping concentration.

Introducing (3) and (4) into (1), the expression of the load line for the MIFIS structure is given as

$$Q_M + Q_{it'} = \frac{C_{TIL}C_{BIL}}{C_{TIL} + C_{BIL}}(V_g - V_{FE} - \varphi_s + \frac{Q_{it'} + Q_{it}}{C_{TIL}}) - Q_{it} \quad (6)$$

According to (5), we find that the distance that the load line shifts up and down depends on the amount of the $Q_{it}$ trapped at the bottom SiO$_x$/Hf$_{0.5}$Zr$_{0.5}$O$_2$ interface, while $Q_{it} + Q_{it'}$ and $C_{TIL}$ determine the distance that the curve shifts left and right.

Based on the principle of the continuity of the electric displacement vector, $Q_M + Q_{it'}$ can be expressed as

$$Q_M + Q_{it'} = C_{FE}V_{FE} + P(V_{FE}) \quad (7)$$

where $C_{FE}$ is the background capacitance of the ferroelectric Hf$_{0.5}$Zr$_{0.5}$O$_2$. The ferroelectric hysteresis loop is obtained by adopting the Preisach model here [36]

$$P(V_{FE}) = P_S \tanh[\alpha(V_{FE} \pm V_C)] \quad (8)$$

where $P_s$ is the saturation polarization, $V_C$ is the coercive voltage, and α is a constant describing how fast the hyperbolic tangent approaches ±$P_s$. Solve (6), (7), and (8) simultaneously, the distribution of voltage and charge of the MIFIS gate stacks at each gate voltage can be obtained by combining the hysteresis loop and the load line.

In simulation, the doping concentration of the silicon substrate is set as 10$^{17}$ cm$^{-3}$. The physical thicknesses of ferroelectric Hf$_{0.5}$Zr$_{0.5}$O$_2$ and bottom SiO$_x$ interlayer are 9.5 nm and 0.7 nm, respectively. The relative dielectric constant of the ferroelectric Hf$_{0.5}$Zr$_{0.5}$O$_2$, top Al$_2$O$_3$ interlayer, and bottom SiO$_x$ interlayer is 30, 9, and 3.9, respectively. For the ferroelectric Hf$_{0.5}$Zr$_{0.5}$O$_2$, the saturated polarization ($P_s$) is set as 23 µC/cm$^2$, and the remanent polarization ($P_r$) is set as 20 µC/cm$^2$. The coercive field is set as 1.5 mV/cm.

IV. RESULTS AND DISCUSSIONS

A. The effect of the top Al$_2$O$_3$ interlayer thickness on the MW

We investigate the dependence of the MW on the pulse amplitude. Fig. 4(a) and (b) show the MW mapping results for the MFIS structure and MIFIS structure with the 5.5 nm top Al$_2$O$_3$ interlayer under the pulse width of 100 µs. For the MFIS structure and MIFIS structure with the 5.5 nm top Al$_2$O$_3$, when the program pulse amplitude goes beyond 5 V or 12 V under the pulse width of 100 µs, respectively, the devices break down. Thus, the maximum MW is 8.4 V for the MIFIS structure with the 5.5 nm top Al$_2$O$_3$ interlayer, while the maximum MW is 1.2 V for the MFIS structure. This indicates that the top Al$_2$O$_3$ interlayer between the metal gate TiN and ferroelectric Hf$_{0.5}$Zr$_{0.5}$O$_2$ can significantly improve the MW. Fig. 5(a) and (b) show the $I_d$-$V_g$ curves corresponding to the maximum MW for the MFIS structure and MIFIS structure with 5.5 nm top Al$_2$O$_3$ interlayer, respectively.

We repeated the above MW mapping measurement process under different pulse widths (not shown here). The maximum MW that can be achieved under different pulse widths is equal to that obtained under the pulse width of 100 µs, as described in [32]. Thus, the pulse width does not affect the above conclusion that the top dielectric between the ferroelectric and the metal gate can significantly increase the MW of the Si-FeFETs.

We study the effect of the top Al$_2$O$_3$ interlayer thickness on the maximum MW. For each sample, we repeated the above MW mapping measurement process. Fig. 6(a) shows the dependence of the maximum MW on the top Al$_2$O$_3$ interlayer thickness. Fig. 6(b) shows the dependence of the $V_{th}$ corresponding to the maximum MW on the top Al$_2$O$_3$ interlayer thickness. We find that the maximum MW first increases at stage I and then remains almost constant at stage II as the thickness of the top Al$_2$O$_3$ interlayer increases. We will discuss its physical origin subsequently.

B. The physical origin of the dependence of maximum MW on the top Al$_2$O$_3$ interlayer thickness

We discuss the physical origin of the dependence of maximum MW on the top Al$_2$O$_3$ thickness in stages I and II,

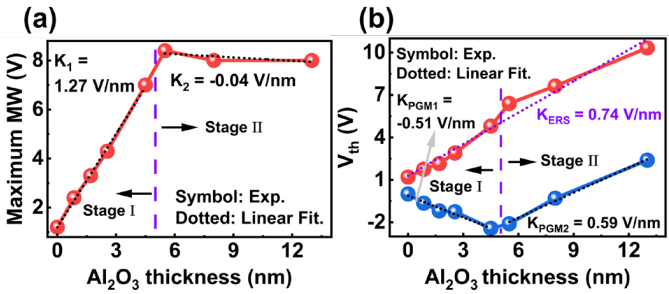

Fig. 6. The dependence of (a) the maximum MW and (b) the $V_{th}$ corresponding to the maximum MW on the top $Al_2O_3$ interlayer thickness.

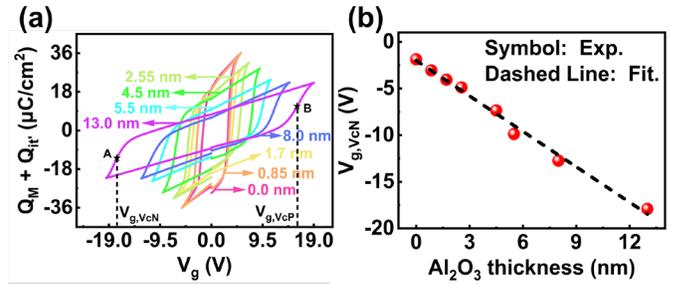

Fig. 7. (a) Results of the PV measurement. (b) The dependence of $V_{g,VcN}$ on the physical thickness of top $Al_2O_3$ interlayer.

respectively. Firstly, we discuss the physical origin of the dependence of the maximum MW on the top $Al_2O_3$ interlayer thickness at stage I as shown in Fig. 6(a). The physical origin of the MW enlargement by inserting the top $Al_2O_3$ interlayer is attributed to the presence of the charges injected from the metal gate trapped at the top $Al_2O_3/Hf_{0.5}Zr_{0.5}O_2$ interface ($Q_{it'}$), as described at [1, 24, 32]. During the program operation, electron de-trapping and/or hole trapping at the top $Al_2O_3/Hf_{0.5}Zr_{0.5}O_2$ interface occurs due to the downward movement of the Fermi energy of the metal gate. After the program operation, positive charges are trapped at the top $Al_2O_3/Hf_{0.5}Zr_{0.5}O_2$ interface because the top $Al_2O_3$ acts as a barrier layer. Then the $V_{th}$ after program operation ($V_{th,PGM}$) negatively shifts. Similarly, after the erase operation, negative charges are trapped at the same interface, causing a positive shift of the $V_{th,ERS}$. The shift of the threshold voltage of the MIFIS structure compared with that of the MFIS structure ($\Delta V_{th}$) is calculated by

$$\Delta V_{th} \approx -\frac{(Q_{it'} + Q_{it})}{\varepsilon_0 \varepsilon_{AlO}} d_{AlO} \quad (9)$$

where $\varepsilon_{AlO}$ and $d_{AlO}$ are the relative dielectric constant and physical thickness of the top $Al_2O_3$ interlayer, respectively. For the MFIS and MIFIS structures, the $Q_{it}$ is identical due to the same bottom $SiO_x$ interlayer. According to (9), the linear dependence of the $V_{th,PGM}$ (or $V_{th,ERS}$) on the top $Al_2O_3$ thickness at stage I indicates that the positive (or negative) charges $Q_{it'}$ after the program (or erase) operation remains constant in the MIFIS structure with different thickness, which is consistent with the description at [1, 35]. Finally, the above results lead to a linear dependence of the maximum MW vs. top $Al_2O_3$ thickness at stage I. Furthermore, the absolute value of the slope of linear dependence corresponding to the program operation is less than that of linear dependence corresponding to the erase operation. This shows that the net number of the positive charges after the program operation is less than the net number of the negative charges after the erase operation according to (9) (the net number of the charges is the sum of the charges trapped at the top $Al_2O_3/Hf_{0.5}Zr_{0.5}O_2$ interface and bottom $SiO_x/Hf_{0.5}Zr_{0.5}O_2$ interface).

Secondly, we discuss the physical origin of the dependence of the maximum MW and $V_{th}$ on the top $Al_2O_3$ interlayer thickness at stage II. As shown in stage I, the $V_{th,ERS}$ still increases linearly with the $Al_2O_3$ interlayer thickness at stage II, which means the $Q_{it'}$ after the erase operation is identical in the MIFIS structure with different top $Al_2O_3$ thicknesses. However, we find that the $V_{th,PGM}$ no longer decreases linearly with the top $Al_2O_3$ interlayer thickness as that in stage I, which results in the maximum MW remaining almost constant with the increasing thickness of the top $Al_2O_3$ interlayer at stage II.

We discuss the physical origin of the fact that $V_{th,PGM}$ does not decrease linearly with the increasing thickness of top $Al_2O_3$ at stage II. According to (9), the dependence of the $V_{th,PGM}$ on the top $Al_2O_3$ thickness at stage II indicates the presence of $Q_{it'} < |Q_{it}|$. Therefore, the reduction in $Q_{it'}$ results in the dependence of the $V_{th,PGM}$ on the top $Al_2O_3$ thickness at stage II. The reduction in $Q_{it'}$ after the program operation may be caused by the two reasons as follows: (i) As the thickness of the top $Al_2O_3$ increases, the hole trapping and/or electron de-trapping at the top $Al_2O_3/Hf_{0.5}Zr_{0.5}O_2$ interface may be more difficult to occur during the program operation due to an increase in the tunneling potential barrier, which may result in the reduction in $Q_{it'}$ during the program operation. (ii) The retention capability of the charges $Q_{it'}$ injected during the program operation degrades rapidly with increasing thickness of the top $Al_2O_3$, leading to a reduction in $Q_{it'}$ before the $V_{th}$ is measured. We discuss their reasonableness in detail as follows.

To investigate whether the charges ($Q_{it'}$) are fully injected during the program operation, we conducted the PV measurement for these devices to obtain the $I_g$-$V_g$ curves. Fig. 7(a) shows the measurement results obtained by integrating the $I_g$. Fig. 7(a) shows that: (i) the slope of the linear part of the total charge density vs. gate voltage ($Q_M+Q_{it}$-$V_g$) curves decreases with increasing thickness of the top $Al_2O_3$, which is attributed to the equivalent gate capacitance decreases with the increasing thickness of top $Al_2O_3$. (ii) The intersections of the $Q_M+Q_{it}$-$V_g$ curves and the $V_g$ axis become closer to the linear part of the $Q_M+Q_{it}$-$V_g$ curves as the thickness of the top $Al_2O_3$ increases. This is caused by a decrease in the measured polarization value. (iii) the $Q_M+Q_{it}$-$V_g$ curves broaden with increasing thickness of the top $Al_2O_3$. Our measurement results are similar to [1]. The physical origin of the broadener of the $Q_M+Q_{it}$-$V_g$ curves is attributed to the partial voltage across the top $Al_2O_3$ interlayer caused by the hysteresis of the charges $Q_{it'}$. Therefore, the degree to which the $Q_M+Q_{it}$-$V_g$ curves broaden can represent the number of charges trapped at the top $Al_2O_3/Hf_{0.5}Zr_{0.5}O_2$ interface. To more reasonably evaluate the degree of broadening of the $Q_M+Q_{it}$-$V_g$ curves, we use the voltage value where the ferroelectric polarization switching rate is 50% to evaluate, such as points A and B shown in Fig. 7(a). Fig. 8(b) shows that gate voltage ($V_{g,VcN}$) corresponding to ferroelectric negative coercive voltage is linearly dependent on the top $Al_2O_3$ interlayer thickness, which indicates the $Q_{it'}$ after the program operation is identical for the MIFIS structure with different top $Al_2O_3$ thicknesses. Therefore, the dependence of the $V_{th,PGM}$ on the top Al$_2$O$_3$ interlayer thickness at stage II is attributed to the reduction in $Q_{it'}$ caused by the retention degradation.

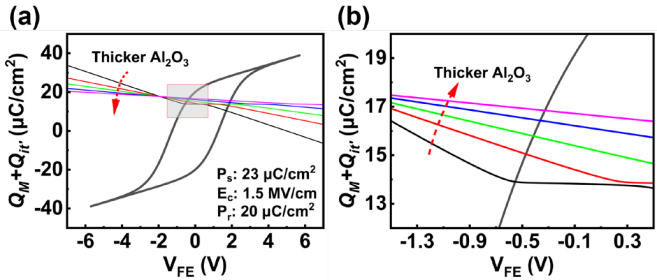

Fig. 8. (a) Hysteresis loop of $Q_M+Q_{it'}-V_{FE}$ and load line at $V_g = 0$ V after the program operation for the different top Al$_2$O$_3$ thicknesses. (b) Enlarged view of the box section in (a).

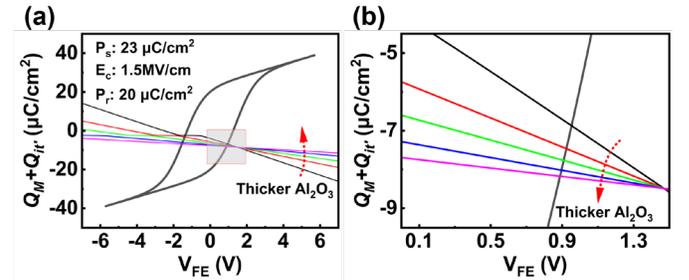

Fig. 10. (a) Hysteresis loop of $Q_M+Q_{it'}-V_{FE}$ and load line at $V_g = 0$ V after the erase operation for the different top Al$_2$O$_3$ thicknesses. (b) Enlarged view of the box section in (a).

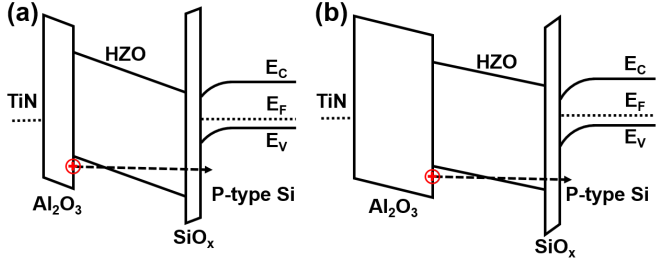

Fig. 9. Energy band diagrams for the MIFIS structure with (a) a thinner Al$_2$O$_3$ and (b) a thicker Al$_2$O$_3$ interlayer at $V_g = 0$ V after the program operation.

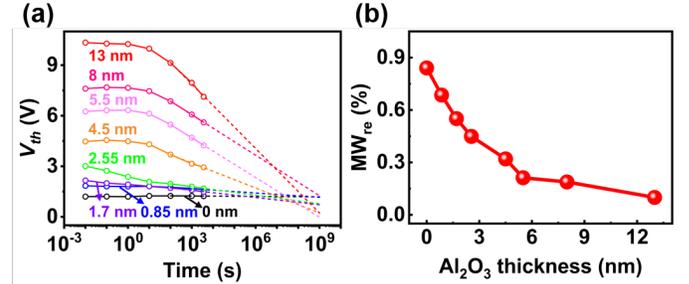

Fig. 11. (a) Retention characteristics after the erase operation for the MIFIS structures with different top Al$_2$O$_3$ thicknesses. (b) The dependence of the MW retention ratio (MW$_{re}$) on the top Al$_2$O$_3$ interlayer thickness.

We use the load line and the ferroelectric hysteresis loop to further explain the physical origin of the retention degradation. We consider the charge trapping phenomenon at the Al$_2$O$_3$/Hf$_{0.5}$Zr$_{0.5}$O$_2$ interface and the bottom SiO$_x$/Hf$_{0.5}$Zr$_{0.5}$O$_2$ interface. Recent research work shows that the charge density trapped at the bottom SiO$_x$/Hf$_{0.5}$Zr$_{0.5}$O$_2$ interface after the program operation is about −13.0 μC/cm$^2$, while the trapped charge density after the erase operation is only about 2.6 μC/cm$^2$ [37]. According to (9) as well as Fig. 6(b), the value $Q_{it'}$ after the program and erase operation is set as 17.7 μC/cm$^2$ and −8.5 μC/cm$^2$, respectively. In simulation, the doping concentration of the silicon substrate is set as $10^{17}$ cm$^{-3}$. The physical thicknesses of ferroelectric Hf$_{0.5}$Zr$_{0.5}$O$_2$ and bottom SiO$_x$ interlayer are 9.5 nm and 0.7 nm, respectively. The relative dielectric constant of the ferroelectric Hf$_{0.5}$Zr$_{0.5}$O$_2$ and top Al$_2$O$_3$ is 30 and 9, respectively. Furthermore, since we are considering the effect of the top Al$_2$O$_3$ interlayer thickness on the maximum MW of the FeFETs with MIFIS structure, i.e., the gate voltage is large enough, we use the saturation hysteresis loop in the simulation.

We conduct the simulation at $V_g = 0$ V after the program operation. Fig. 8(a-b) shows the hysteresis loop of $Q_M+Q_{it'}-V_{FE}$ and load line at $V_g = 0$ V after the program operation for different top Al$_2$O$_3$ thicknesses. We find that the electric field of the ferroelectric Hf$_{0.5}$Zr$_{0.5}$O$_2$ gradually decreases with increasing thickness of the top Al$_2$O$_3$ as shown in the Fig. 8(b). This means that the charges $Q_{it'}$ face a smaller tunneling potential barrier for the MIFIS structure with a thicker top Al$_2$O$_3$ when $Q_{it'}$ tunnel to the Si substrate, as shown in Fig. 9, which results in more charges $Q_{it'}$ tunneling to the Si substrate to recombine with electrons. The above result leads to a decrease in $Q_{it'}$, so that $Q_{it'}$ becomes less than $|Q_{it}|$ ($Q_{it'} < |Q_{it}|$) after the program operation in the MIFIS structure with a thicker top Al$_2$O$_3$ interlayer. Therefore, the $V_{th,PGM}$ no longer decreases linearly with increasing thickness of the top Al$_2$O$_3$ interlayer at stage II.

In addition, we perform the simulation at $V_g = 0$ V after the erase operation to further discuss the reason why the charge trapped at the top Al$_2$O$_3$/Hf$_{0.5}$Zr$_{0.5}$O$_2$ interface ($Q_{it'}$) after the erase operation can be maintained. Fig. 10 shows the hysteresis loop of $Q_M + Q_{it'} - V_{FE}$ and load line at $V_g = 0$ V after the erase operation for the different top Al$_2$O$_3$ thicknesses. According to (6), we find that compared with the program operation, the load line shifts downward by a smaller distance due to the smaller charge density trapped at the bottom SiO$_x$/Hf$_{0.5}$Zr$_{0.5}$O$_2$ interface ($Q_{it}$) after the erase operation. Therefore, the change in the electric field of the ferroelectric Hf$_{0.5}$Zr$_{0.5}$O$_2$ after the erase operation caused by the change in thickness of the top Al$_2$O$_3$ interlayer is reduced. This indicates that there is only a slight difference in tunneling potential barriers faced by $Q_{it'}$ in the MIFIS structure with different Al$_2$O$_3$ thicknesses. The slight difference does not cause a significant decrease in $Q_{it'}$ within a short time. Thus, the $V_{th,ERS}$ is linearly dependent on the thickness of the top Al$_2$O$_3$ interlayer over the entire thickness range.

### C. The effect of top Al$_2$O$_3$ thickness on the endurance and retention characteristics

We investigate the dependence of the retention characteristics under the pulse amplitude corresponding to the maximum MW on top Al$_2$O$_3$ interlayer thicknesses. Fig. 11(a) shows the retention characteristic after the erase operation for the MIFIS structures with different top Al$_2$O$_3$ interlayer thicknesses. We find that the $V_{th,ERS}$ first remains almost constant within a time range of 10 s and then decreases with an increase in time. In addition, this degradation rate of $V_{th,ERS}$ increases with the increasing thickness of the top $Al_2O_3$ interlayer.

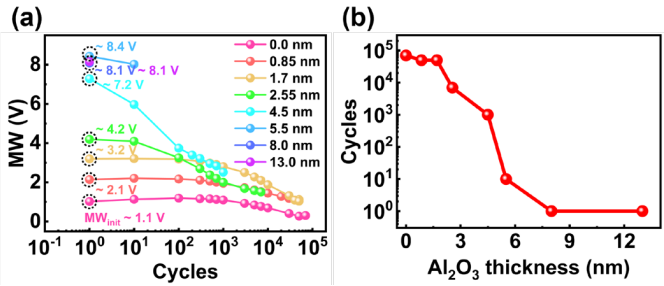

Fig. 12. (a) Endurance characteristics corresponding to the maximum MW. (b) The dependence of the cycles that correspond to the device endurance failure on the top $Al_2O_3$ interlayer thickness.

We discuss the physical origin of the dependence of the retention characteristics after the erase operation on top $Al_2O_3$ thickness. As the above discussion, the electric field of the ferroelectric $Hf_{0.5}Zr_{0.5}O_2$ after the erase operation decreases as the thickness of the top $Al_2O_3$ increases, as shown in Fig. 10. This leads to the fact that the degradation rate of the $V_{th,ERS}$ increases with the increasing thickness of the top $Al_2O_3$ interlayer. In addition, on the whole, the electric field of the ferroelectric $Hf_{0.5}Zr_{0.5}O_2$ only exhibits a slight difference for the different thicknesses of the top $Al_2O_3$. The slight difference in the electric field results in a slight difference in tunneling potential barriers. Nevertheless, the slight difference of tunneling potential barriers does not cause a significant decrease in $Q_{it'}$ within a short time, Thus, the $V_{th,ERS}$ remains almost constant within a time range of 10 s for the different top $Al_2O_3$ interlayer thicknesses.

We define the ratio of the $MW_{end}$ (delay = $3.15 \times 10^8$ s) to the $MW_{init}$ (delay = 10 ms) as the MW retention ratio ($MW_{re}$), i.e.,

$$MW_{re} = \frac{MW_{end}}{MW_{init}} \quad (9)$$

Fig. 11(b) shows the dependence of the $MW_{re}$ on the top $Al_2O_3$ interlayer thickness. We find the $MW_{re}$ decreases with the increasing thickness of the top $Al_2O_3$ interlayer. This indicates the MIFIS structure with a thicker top $Al_2O_3$ has poorer retention characteristics. The retention characteristics after the program operation exhibit little difference compared with those of the erase operation (not shown here), and therefore, the above phenomenon is attributed to the dependence of the degradation rate of the $V_{th,ERS}$ on the thickness of the top $Al_2O_3$ interlayer.

Furthermore, we study the impact of top $Al_2O_3$ interlayer thickness on the endurance characteristics under the pulse amplitude of the corresponding maximum MW. Fig. 13(a) shows the measurement results of endurance characteristics corresponding to the maximum MW. We define the cycles corresponding to MW = 0 or breakdown failure of gate stacks as the endurance failure cycles of the device. Fig. 13(b) shows the dependence of the endurance failure cycles on the top $Al_2O_3$ interlayer thickness. We find the MIFIS with a thicker top $Al_2O_3$ dielectric interlayer has poorer endurance characteristics, which is similar to the degradation of the endurance characteristics with the increasing thickness of the top $SiO_2$ interlayer described in [35]. The possible physical origin is the breakdown of the top $Al_2O_3$ dielectric interlayer. Since the rate of ferroelectric polarization switching is higher than the rate of charge trapping and de-trapping [14, 37-39]. Therefore, the ferroelectric polarization switches in a shorter time after the application of the program pulse, and the large spontaneous polarization ($P_s$) of ferroelectric $Hf_{0.5}Zr_{0.5}O_2$ leads to an electric field of the top $Al_2O_3$ that is almost close to its breakdown electric field. However, as the thickness of the top $Al_2O_3$ interlayer increases, the time that the same amount of charge is injected from the metal gate to the top $Al_2O_3/Hf_{0.5}Zr_{0.5}O_2$ interface ($Q_{it'}$) increases. This indicates that the thicker top $Al_2O_3$ dielectric interlayer experiences a longer time at a high electric field. In addition, some studies have shown that the dielectric breakdown strength ($E_{bd}$) reduces as the dielectric thickness increases [40]. Thus, the MIFIS structure with a thicker top $Al_2O_3$ interlayer is more susceptible to breakdown under pulse amplitude corresponding to maximum MW. This leads to the phenomenon that the endurance characteristics deteriorate with increasing thickness of the top $Al_2O_3$ interlayer.

## V. CONCLUSIONS

In our work, we study the effect of the top $Al_2O_3$ interlayer thickness on the MW of FeFETs with the MIFIS gate structure. We find the MW first increases and then remains almost constant with the increasing thickness of the top $Al_2O_3$. The phenomenon is attributed to the lower electric field of the ferroelectric $Hf_{0.5}Zr_{0.5}O_2$ in the MIFIS structure with a thicker top $Al_2O_3$ after a program operation, which makes the charges ($Q_{it'}$) trapped at the top $Al_2O_3/Hf_{0.5}Zr_{0.5}O_2$ interface cannot be retained. Furthermore, we investigate the effect of the top $Al_2O_3$ interlayer thickness on the reliability (endurance characteristics and retention characteristics). We find that the MIFIS structure with a thicker top $Al_2O_3$ interlayer has poorer retention endurance characteristics.